\documentclass[prb,aps,twocolumn,showpacs,floatfix]{revtex4}

\usepackage{epsfig}

\begin{document}

\title{Comment 
on ``Accurate ground-state phase diagram of the 
one-dimensional extended Hubbard model at half filling''}

\author{Eric Jeckelmann} 
\affiliation{Institut f\"{u}r Physik, Johannes 
Gutenberg-Universit\"{a}t, 55099 Mainz, Germany}

\date{\today}

\begin{abstract}
It is shown that Guoping Zhang's results
[G.P. Zhang, \prb \textbf{68}, 153101 (2003)]
for the charge-density-wave
phase boundary in the half-filled one-dimensional extended
Hubbard model are incorrect
and that his criticism of my work
[E. Jeckelmann, \prl \textbf{89}, 236401 (2002)]
is groundless.
\end{abstract}


\maketitle

In Ref.~\onlinecite{Zhang}, Guoping Zhang presented density-matrix 
renormalization group (DMRG) results 
which contradict my DMRG 
calculations~\cite{Jeckelmann} and Hirsch's quantum Monte Carlo
(QMC) simulations~\cite{Hirsch} for 
the charge-density-wave 
(CDW) phase boundary in the one-dimensional extended
Hubbard model at half filling. 
In this Comment I show that Guoping Zhang's results are inaccurate
and that his criticism of my work is groundless.

\begin{figure}[b]
\begin{center}
\epsfig{file=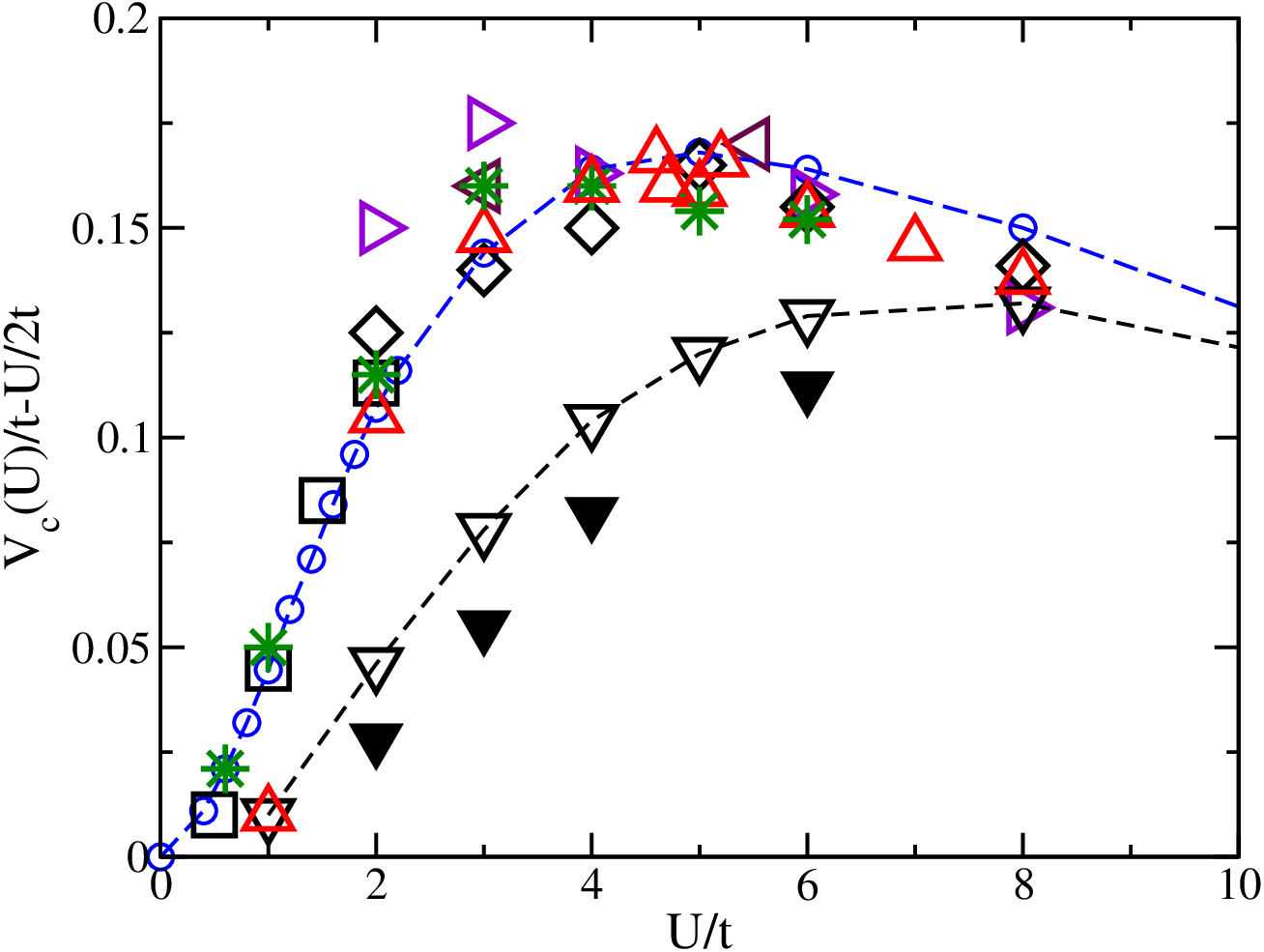,width=7.6cm}
\end{center}
\caption{(Color online) Results for the CDW phase boundary $V_c(U)$:
QMC simulations (Ref.~\onlinecite{Hirsch}) (right triangle), 
exact diagonalizations (Ref.~\onlinecite{Cannon}) (left triangle), 
level crossing analysis (Ref.~\onlinecite{Nakamura}) (circle),
SSE-QMC simulations (Refs.~\onlinecite{Sengupta} and~\onlinecite{Sandvik}) 
(up triangle), 
author's DMRG calculations 
[from Ref.~\onlinecite{Jeckelmann} 
(diamond) and new results (square)], 
Yuzhong Zhang's DMRG calculations 
(Ref.~\onlinecite{YZZhang}) (star),
and Guoping~Zhang's DMRG calculations (infinite-system algorithm)
in Ref.~\onlinecite{Zhang} (open down triangle)
and in a previous work (Ref.~\onlinecite{oldZhang}) (solid triangle).  
Dashed lines are guides for the eye.}
\label{fig1}
\end{figure}

Although the phase diagram of the extended Hubbard model
is still partially controversial (see
Refs.~\onlinecite{Jeckelmann,Sandvik,YZZhang} and references therein),
the CDW phase boundary $V_c(U)$ in the parameter space ($U,V$)
was determined years ago~\cite{Hirsch,Cannon} and has not
been disputed in recent 
studies.~\cite{Jeckelmann,Sandvik,YZZhang,Sengupta,Nakamura}
In Fig.~\ref{fig1}, I show the results of various 
numerical investigations
for $V_c(U)-U/2$ in the weak to intermediate coupling regime.  
There is an excellent overall agreement between
Hirsch's QMC simulations,~\cite{Hirsch} 
the exact diagonalizations of
Cannon \textit{et al.},~\cite{Cannon} Nakamura's 
level crossing analysis,~\cite{Nakamura} the stochastic
series expansion QMC (SSE-QMC) simulations of
Sandvik \textit{et al.},~\cite{Sengupta,Sandvik}
Yuzhong Zhang's DMRG calculations~\cite{YZZhang}, 
and my DMRG calculations~\cite{Jeckelmann}. 
In particular, my results agree quantitatively~\cite{errors}
with the most recent and accurate 
numerical simulations.~\cite{Nakamura,Sengupta,Sandvik,YZZhang}
Only Guoping Zhang's DMRG data~\cite{Zhang,oldZhang}
deviate systematically from the other results. 
Therefore, there is clearly a problem with his calculations.

\begin{figure}[t]
\begin{center}
\epsfig{file=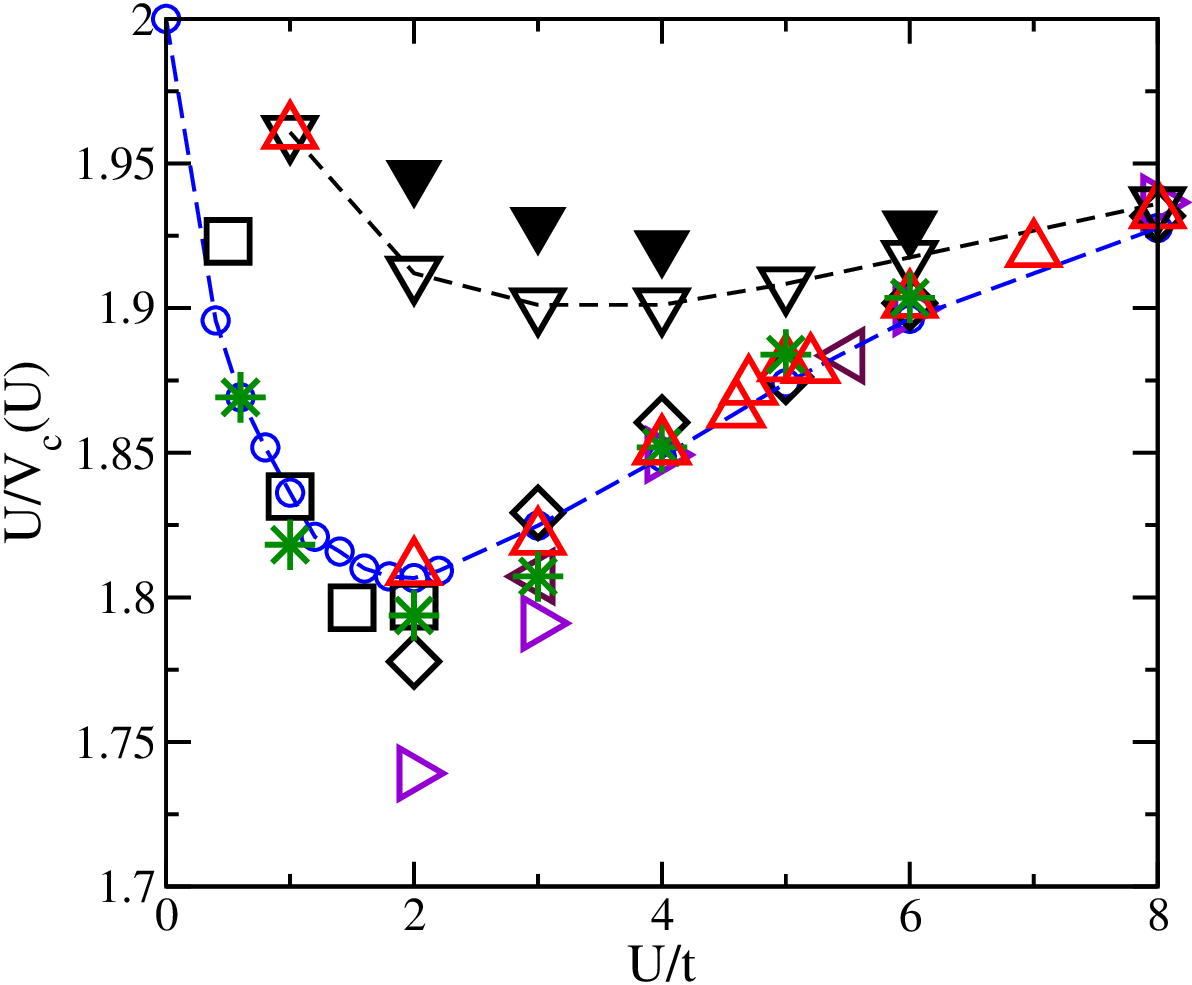,width=7.6cm}
\end{center}
\caption{(Color online) Same results as in Fig. 1 but displayed
using Zhang's representation ($U/V_c$ vs $U$).
}
\label{fig2}
\end{figure}

The discrepancy between the various
DMRG calculations~\cite{Zhang,Jeckelmann,YZZhang,oldZhang}
is not surprising. 
Guoping Zhang uses the \textit{infinite-system}
DMRG algorithm while Yuzhong Zhang
and I use the more accurate \textit{finite-system}
DMRG algorithm.~\cite{White}
It is well known~\cite{Schollwoeck}
that for many problems 
the infinite-system algorithm yield incorrect
results while the finite-system algorithm
gives essentially exact (numerical) results.  
In particular, 
it is essential to use the more reliable finite-system DMRG algorithm
for inhomogeneous systems such as a CDW ground state. 
Therefore, the discrepancy between Guoping Zhang's results
and all other works just demonstrates the failure of
the standard infinite-system DMRG algorithm for the present problem.
(See Ref.~\onlinecite{Prokofiev} for another example  
of the infinite-system algorithm failure and Ref.~\onlinecite{Rapsch}
for the successful investigation of the same problem with
the finite-system algorithm.)

In his paper Guoping Zhang wrongly claimed that my DMRG calculations 
(and the
QMC simulations of Ref.~\onlinecite{Hirsch})
failed to reproduce the weak coupling limit result $V_c(U) = U/2$.
In Refs.~\onlinecite{Jeckelmann,Hirsch} the investigation of the phase
diagram was focused on the
intermediate- and strong-coupling regimes (i.e., $U \geq 2t$)
and no analysis of the weak-coupling limit $U \ll t$
was performed. 
Here I present additional results for $V_c(U)$ calculated with 
DMRG for weaker couplings: 
$V_c/t = 0.260 \pm 0.003$ for $U/t=0.5$,
$V_c/t=0.545 \pm 0.005$ for $U/t=1$, and 
$V_c/t=0.835 \pm 0.005$ for $U=1.5t$.
Moreover, I have calculated $V_c(U=2t)/t$ more accurately
and found $1.113 \pm 0.005$ (in agreement within the
error bars with the 
value given in Ref.~\onlinecite{Jeckelmann}).
These results are shown in Fig.~\ref{fig1} as square.
They agree perfectly with other works~\cite{Sandvik,YZZhang,Nakamura}
and, clearly,
they approach the weak-coupling result $V_c(U)=U/2$
in the limit $U \rightarrow 0$.
In Figs.~1 and~2 of Ref.~\onlinecite{Zhang} Guoping Zhang used
a different representation of the data,
$U/V_c(U)$ vs $U$, to analyze the weak-coupling limit.
In Fig.~\ref{fig2}, I show again all data of
Fig.~\ref{fig1} using this representation.
Clearly, the minimum of the ratio $U/V_c(U)$ 
occurs for $U$ slightly smaller than $2t$
and the weak-coupling limit is
recovered only for $U$ smaller than $2t$.
Therefore, 
the $U \rightarrow 0$ limit of $U/V_c(U)$ cannot be determined
using numerical data for $U \geq 2t$ and
the Figs.~1 and~2 of Ref.~\onlinecite{Zhang} are
misleading. 
The DMRG and QMC data for $U \geq 2t$ presented in 
Refs.~\onlinecite{Jeckelmann} and~\onlinecite{Hirsch}
are fully compatible with the weak-coupling limit $V_c(U) = U/2$,
contrary to Guoping Zhang's assertion in Ref.~\onlinecite{Zhang}.

In summary, comparisons with the results available in the
literature confirm the accuracy and reliability
of the DMRG calculations presented in Ref.~\onlinecite{Jeckelmann}. 
Guoping Zhang's results and conclusion are faulty due to the 
inappropriate use 
of the infinite-system DMRG algorithm.


\begin{thebibliography}{99}
\bibitem{errors} The error margins 
for $V_c(U)$ in 
Refs.~\onlinecite{Sandvik,YZZhang,Sengupta}
are smaller than $0.01t$ (and thus smaller than the 
corresponding symbols in Fig.~\ref{fig1}), 
excepted for the SSE-QMC point at $U=t$, which has
rather large but not precisely quantifiable error 
bars.~\cite{Sandvik}
\end{thebibliography}
\end{document}